\documentclass[twocolumn,showpacs,preprintnumbers,amsmath,amssymb,superscriptaddress]{revtex4}

\usepackage{graphicx}
\usepackage{dcolumn}
\usepackage{bm}
\usepackage{longtable}

\begin{document}
\title{Interacting Random Walkers and Non-Equilibrium Fluctuations}

\author{E. Agliari}
\affiliation{Dipartimento di Fisica, Universit\`a degli Studi di
Parma, viale Usberti 7/A, 43100 Parma, Italy}
\affiliation{Theoretische Polymerphysik, Universit\"{a}t Freiburg, Hermann-Herder-Str. 3, D-79104 Freiburg, Germany}
\author{M. Casartelli}
\affiliation{Dipartimento di Fisica, Universit\`a degli Studi di
Parma, viale Usberti 7/A, 43100 Parma, Italy}
\affiliation{CNR-INFM, Gruppo Collegato di Parma, viale Usberti
7/A, 43100 Parma, Italy} \affiliation{INFN, Gruppo Collegato di
Parma, viale Usberti 7/A, 43100 Parma, Italy}
\author{A. Vezzani}
\affiliation{Dipartimento di Fisica, Universit\`a degli Studi di
Parma, viale Usberti 7/A, 43100 Parma, Italy}
\affiliation{CNR-INFM, Gruppo Collegato di Parma, viale Usberti
7/A, 43100 Parma, Italy}

\date{\today}

%
\begin{abstract}
We introduce a model of interacting Random Walk,
whose hopping amplitude depends on the number
of walkers/particles on the link. The mesoscopic counterpart of such a
microscopic dynamics is a diffusing system whose diffusivity
depends on the particle density. A non-equilibrium stationary flux
can be induced by suitable boundary conditions, and we show indeed
that it is mesoscopically described by a Fourier equation with a
density dependent diffusivity. A simple mean-field description
predicts a critical diffusivity if the hopping amplitude vanishes
for a certain walker density. Actually, we evidence
that, even if the density equals this pseudo-critical value, the
system does not present any criticality but only a dynamical
slowing down. This property is confirmed by the fact that, in
spite of interaction, the particle distribution at equilibrium
is simply described in terms of a product of Poissonians.
For mesoscopic systems with a stationary flux,
a very effect of interaction among particles consists
in the amplification of fluctuations, which is especially
relevant close to the pseudo-critical density.
This agrees with analogous results obtained
for Ising models, clarifying that larger fluctuations
are induced by the dynamical slowing down and not by a genuine criticality.
The consistency of this amplification effect with altered coloured noise
in time series is also proved.\end{abstract}


\maketitle

\section{Introduction}
\label{intro}

Diffusive dynamics is at the core of the study of heat or matter
transport \cite{mehrer,karger}. In fact, exceptions to this
paradigm are themselves matter of research \cite{livi}. Typically,
one expects that non-equilibrium steady states are described by a
Fourier-like equation with non constant diffusivity; for
homogeneous systems, such a diffusivity should depend on the local
density of energy or matter. Actually, this kind of behaviour was
revealed in different fields - magnetic systems, biology,
reaction-diffusion processes (see for example
\cite{exper,pawlak,flierl,kampen}). On a more modellistic ground,
we may quote a recently studied 2-dimensional Ising system between
two thermostats, where  heat transport exhibits indeed an energy
dependent diffusivity \cite{harris,saito,cmv,acv}.

A basic question is the relation between the mesoscopic
description given by the Fourier equation and the underlying
microscopic dynamics. In this context, geometry, interactions and possible
criticalities should also be taken into account. In fact, various
features depend on this mesoscopic-microscopic relation, e.g.
scaling properties, behaviour of fluctuations, coexistence of
critical and non-critical phases within steady states, etc.

In order to carry out the analysis, a suitable resource is given
by the prototype of diffusive discrete processes, i.e. random
walks (RW) \cite{weiss}. In its original form, a particle (the
walker) has a constant uniform probability to jump at discrete
times from a site to another in its neighborhood. The addition of
constraints (e.g. self avoiding rules) or the dependence of the
probability on the site or the neighborhood state can fit a
variety of interesting situations, including non-equilibrium
dynamics. Moreover, particle reservoirs with different densities
may be fixed at borders, determining a flux from the
``hot'' (dense) to the ``cold'' (rare)
side. Therefore, we expect that interacting RW could represent
various kinds of transport properties. In this way, the particle
population can be read  as a diffusing ``energy''.
Such an interpretation fits the behaviour of
many diffusive systems, physically very different in principle,
from sandpiles to spin models, as those quoted above.

A number of conceptual achievements have been obtained in these fields
focusing in particular on the symmetric simple exclusion process (SSEP),
where two walkers cannot occupy the same site,
and  on the zero-range processes (ZRP) where the
hop rate depends on the particle density of
the departure site (see reviews \cite{evans,schutz}).
An obvious extension of ZRP are the
misanthrope processes (MP), where the hop rate depends also on the
density of the target site. Actually, MP have been  investigated
both for the factorization conditions of equilibrium states and for
the mapping onto other models. Moreover, for SSEP and ZRP the
situation where a stationary flow is imposed by the presence of
particle reservoirs has also been considered.
In particular, the problems of the stationary density profile \cite{lebowitz}
and of current fluctuations \cite{derrida,rakos,bodineau}
have been deeply investigated.

Here, we want to focus on
the possibility of modelling some of the features characterizing a
diffusive dynamics with non constant diffusivity. We would give evidence,
for instance, to the relevance of a possible critical
phase in mesoscopic phenomena, especially those regarding
fluctuations and decay of correlations.

Consider for definiteness a one dimensional discrete array with particle
reservoirs at the extremes (the generalization to higher
dimensional lattices is straightforward). We
shall introduce a particular case of symmetric MP where the hop probability
is ``totalistic'': this means that it depends only on the mean value,
hereafter called ``link-population'',  between the departure-arrival
populations. Such a model is expected to describe a
general diffusive system where the local conductibility of a
microscopic connection depends on the average density of the link
itself, i.e where diffusivity is density dependent. Moreover,
for a possible critical behaviour, the dynamics
should also include a parameter $N_c$, such that the hop rate
tends to 0 as the link population approaches $N_c$. In a
mean field description, this implies a vanishing diffusivity
at $N_c$, suggesting indeed a possible criticality.

Analytical and numerical investigations show that our RW
system actually exhibits diffusion, being well described
by a Fourier equation for the population density. In
particular, in agreement with our intent, the diffusivity depends
on the average population. Moreover, there is a dynamical slowing
down when this population equals $N_c$, and mesoscopic realizations
prove that there exists a band around this putative critical parameter
where fluctuations are amplified with respect to the
equilibrium values. However, it is possible to show that $N_c$ is
not a true critical point; for large sizes, in particular,
fluctuations tend to vanish in a clear analytical way.

This constitutes an interesting result, especially in comparison
with the spin system mentioned above \cite{cmv,acv}. There are
relevant analogies: the spin flip in Ising and the jump
probability in RW both depend on ``~link properties~'' (energy and
link populations, respectively); both systems display a
pseudo-critical parameter (the critical temperature or energy, say $E_c$,
and the population $N_c$, respectively). Nonetheless, since
evolution rules are different, there are also
important differences: not only $E_c$ and $N_c$ have a different
critical nature, but also dimensionality, for instance, plays a
quite different role.
Since for both systems we evidence amplified fluctuations at the
mesoscopic scale, we can conclude that their source is not the
(virtual) presence of a genuine critical point, but rather the
slowing down in dynamics induced by particle interaction.

After the introduction of model and formalism (section
\ref{sec:stationary}), we investigate the system first by
analytical methods and then by numerical simulations. The
latter corroborate and expand analytical results.
More precisely, in section \ref{sec:factorization} we prove that,
for a closed system, a hopping probability depending on the link
population is sufficient to derive a factorized Poissonian
distribution describing the statistics of the local population.
Then, in Section \ref{sec:diffusivity}, we take advantage of this
result to get an analytic expression for the diffusivity, which is
valid for sufficiently large systems. Such an expression is confirmed by
numerical simulations displayed in Section \ref{sec:numerical}.
We also discuss data relevant to fluctuations
for proper observables and their possible correlations with the
quasi-critical region, the meaning of such a dynamical slowing
down and the mesoscopic character of the phenomenon. Finally,
we study internal correlations by means of the spectral analysis
of time series.

\section{The model}
\label{sec:stationary}
A random walk is a stochastic process defined on a generic graph.
At each discrete time-step, the walker moves to an adjacent site
following a random rule which is independent of the walk history,
so that the process is Markovian. For the simple random walk,
the walker jumps on one of its neighbouring sites with uniform probability.

In a system of $N$ simple walkers (particles) on an Euclidean lattice
the limit of continuous time and space is described by the
Fourier-like diffusion equation
\begin{equation}
{\partial}_t \mu({\mathbf x},t) = - \overrightarrow{\partial}\cdot
(D \overrightarrow{\partial} \mu ({\mathbf x},t)),
\label{fourier}
\end{equation}
where $\mu(\mathbf{x},t)$ is the particle density in $\mathbf{x}$
at time $t$ and the constant $D$ is the diffusivity. Note that the
quantity which is derived on the right side of (\ref{fourier}),
i.e.
\begin{equation} \label{eq:flux_fourier}
\phi(x)= - (D \overrightarrow{\partial} \mu ({\mathbf x},t)),
\end{equation} is the local flux at $x$.

Equation (\ref{fourier}) has been introduced in different contexts
to study matter or energy transport. In particular, even for
several interacting models, transport is expected to be described
by an analogous Fourier-like equation with non constant
diffusivity \cite{mehrer,karger}. In this work, we aim to
introduce a simple microscopic model where, at mesoscopic level, the
diffusivity depends on the particle density $\mu$ in a tunable
way.

More precisely, we consider a system of walkers on a one dimensional
lattice of $L$ sites $i=1,...,L$
(anyway, most of our results can be generalized straightforwardly
to higher dimensional lattices).
We study both the close chain with periodic boundaries and
the system with open boundaries in contact with particle reservoirs.
The transition probability between neighbouring sites is
a symmetric function of $n_i$ and $n_{i+1}$,
i.e. the number of walkers occupying sites $i$ and $i+1$ respectively. In
the following, $n_i$ will be called occupation number or site
energy, and  $E_{i}=(n_i+n_{i+1})/{2}$ will be called link population or
link energy. In order to simplify formulas, we shall often use
$\mu_i$ instead of the more explicit $\langle n_i \rangle $.

 Now, let $f(n_i,n_{i+1})$ be the probability for a particle
on the site $i$ to jump on the adjacent site ${i+1}$.

In a mean field approach, we discard the number fluctuations,
assuming that $n_i\simeq \langle n_i \rangle$. The average number
of particles crossing the link  $i, {i+1}$ in
the unit time, i.e. the particle flux, is
$$
\Phi(i,i+1) = \langle n_i \rangle f(\langle n_i\rangle ,
\langle n_{i+1}\rangle) - \langle n_{i+1} \rangle f(\langle n_{i+1}\rangle ,\langle n_i \rangle).
$$
If the transition probability $f(n_i,n_{i+1})$
is symmetric in its arguments, we can write:
\begin{eqnarray}
\label{eq:flux}
\Phi(i,i+1)& = & (\langle n_i \rangle -\langle n_{i+1} \rangle )
f(\langle n_{i+1}\rangle ,\langle n_i \rangle)\nonumber\\
& = & - \frac{\partial \langle n\rangle}{\partial x} f(\langle n_{i+1}
\rangle,\langle n_i\rangle)
\end{eqnarray}
where  ${\partial \langle n \rangle}/{\partial x}$, or equivalently
${\partial \mu }/{\partial x}$, represents the discrete
spatial derivative for the average number of particles.

The previous expression simply states that the flow is
proportional to the particle gradient. Therefore, it is a
discrete one dimensional version of the flux
(\ref{eq:flux_fourier}) appearing in the Fourier equation
(\ref{fourier}).
 Now, if $f(n_{i+1},n_i)$ depends only on the link energy
$E_i$, and if the system is large enough so that the energy
differences between neighbouring sites can be neglected, by comparing
Eq.~\ref{eq:flux} and Eq.~\ref{fourier}, we can express the
diffusivity as $D(\mu)=f(\mu)$.

It must be underlined that the approach followed in this section
is a mean-field type, since fluctuations in particle density have
not been taken into account. In the next sections we make a step
forward by introducing a probability distribution for the
particle configurations, and we obtain a  more reliable expression
for the diffusivity.

Finally, we notice that if Eq.~\ref{fourier} holds, the
diffusivity $ D(\mu)$ can be calculated directly from the spatial
distribution of the average density $\mu_i$. In fact, in a stationary
system, the average flux is constant along the direction of the underlying
structure. Hence we obtain
\begin{equation}
\label{eq:stationary_flow} D(\mu) = C
\left ( \frac{\partial \mu}{\partial x} \right)^{-1},
\end{equation}
where the constant $C$ is the flux.

If for different system
parameters (size, boundary conditions,...) Equation
(\ref{eq:stationary_flow}) provides the same expression of the
diffusivity as a function of the density, this would be a reliable
check of the validity of the Fourier equation (\ref{fourier}) in
describing matter transport in the system \cite{cmv}.

\section{Analytical Results on close systems}
\label{sec:factorization}

Consider a close system of $N$ particles
with the following updating rule: at each time step,
a randomly chosen particle can move
from its site $i$, towards sites ${i+1}$ or ${i-1}$ , or else stop,
according to the transition probabilities $f_r = f(n_i,n_{i+1})$, $f_l =
f(n_i,n_{i-1})$ and $f_s = 1 - f_r - f_l$, respectively.
Hence, when the particle moves from $i$ to, say, ${i+1}$, then $n_i$ changes into $n_{i}'=n_{i}-1$ while $n_{i+1}$ changes into $n_{i+1}'=n_{i+1}+1$.

The stationary state of such a system is described by the
distribution $P(\{ n \})$, where $\{ n \}$ represents the
particle configuration defined by the occupation numbers of each site.
Obviously, $\sum_i n_i = N$.

The detailed balance condition reads:
\begin{equation}
\label{eq:detailed} P(\{ n \} ) \frac{n_i}{N} f(n_i,n_{i+1}) =
P(\{ n' \} ) \frac{n_{i+1}'}{N} f(n_{i+1}',n_i'),
\end{equation}
where $\frac{n_i}{N}$ represents the probability that one of the particles
on $i$ is selected. The solution of this equation would
provide the probability distribution for the equilibrium system.

As already noticed, if a transition probability is totalistic, i.e.
it depends on the sum $n_i + n_{i+1}$, then $f(n_i,n_{i+1})=
f(n_{i+1}',n_i')$. In fact, $n_{i+1}'+n_{i}'=n_{i+1}+1+n_i-1$.
Therefore, Eq.~\ref{eq:detailed} gets
\begin{equation}
\label{eq:detailed_simpl}
P(\{ n \}) n_i  = P(\{n'\}) n_{i+1}'.
\end{equation}

We note that totalistic transition probabilities satisfy the condition given in
reference \cite{evans} for the factorizability of the distribution function in a MP.
Therefore we can assume:
\begin{equation}
\label{eq:factor}
P(\{ n \}) = \prod_{i} P_i(n_i).
\end{equation}
The previous equation allows to focus, in Eq.~\ref{eq:detailed_simpl},
on the factors relevant to indices $i$ and $i+1$ only, giving
\begin{equation}
\label{eq:detailed_simplsimpl} P_i(n_i) P_{i+1}(n_{i+1}) n_i  =
P_i(n_i-1) P_{i+1}(n_{i+1}+1) n_{i+1}.
\end{equation}
By properly separating terms we obtain the following recursive
equation:
\begin{equation}
\label{eq:recursive} \frac{P_i(n_i)}{P_i(n_i-1)} n_i =
\frac {P_{i+1}(n_{i+1}+1)} {P_{i+1}(n_{i+1})} n_{i+1}= \mathrm{const},
\end{equation}
whose solution is just a Poissonian distribution independent of
the site index (the constant does not depend neither on the site nor
on the particle number).
More precisely, we have
\begin{equation}
P_{i}(n_i)=\label{eq:poisson} \mathcal{P}_{\mu}(n_i) = \frac{e^{-\mu}
\mu^{n_i}}{n_i!}
\end{equation}
where $\mu$ is the average number of particles per site.

Clearly, in the presence of a density gradient, the previous
expression is no longer correct. In fact, the mean occupation
number $\mu$ is in principle site dependent. However, we expect
that the Poissonian distribution still provides a good
approximation when the system is sufficiently large.

Finally, it is important to notice that the request imposed on the transition
probability $f$  (i.e. to be dependent on the sum $ n_i+n_{i+1}$ only)
proved to be sufficient to get the distribution (\ref{eq:poisson}).
We remark also that the equilibrium distribution depends on the particle
density but not on the form of $f(n_{i+1},n_i)$.
The Poissonian distribution therefore characterizes a wide class of random
walk systems, and not only the simple non-interacting model.

\section{Analytical approach to diffusivity} \label{sec:diffusivity}

Let us now consider a system endowed with a density gradient: two
particle reservoirs are placed at the boundaries forcing
the occupation numbers of border sites to assume different
values. More precisely, at each time step the particle numbers on sites
$i=1$ and $i=L$ are drawn from Poissonian distributions with averages $N_1$ and
$N_L$ respectively ($N_1<N_L$).

The flux between sites $i$ and $i+1$ (Eq.~\ref{eq:flux})
has to be rewritten taking into account the
fluctuations of the particle numbers:
\begin{eqnarray}
\Phi(i,i+1) & = &
\sum_{\{n \}} {P}(\{n \}) n_i f(n_{i}+n_{i+1}) -
\nonumber\\
& & -
 \sum_{\{n \}} {P}(\{n \}) n_{i+1} f(n_{i}+n_{i+1})
\label{eq:flux_poisson}
\end{eqnarray}
The previous expression is exact, since it takes into account all
possible configurations. Conversely, Eq.~\ref{eq:flux},
where occupation numbers are treated as average values,
is a mean-field approximation of Eq.~\ref{eq:flux_poisson}.

Now, in the presence of a density gradient,
the analytical expression for the configuration probability ${P}(\{n \})$
is not known. However, if we consider a large system such that $N_L - N_1 \ll L $,
a reasonable ansatz is to assume
that the distribution is still factorizable and that in each site
the occupation probability is described by a site dependent Poissonian.

The average occupation number of site $i$ will be
provisionally denoted $\mu$ instead of $\mu_i $.
Assuming a certain smoothness, for the adjacent
site we can consistently write
$\mu_{i+1} = \mu + \varepsilon$, with $\varepsilon $ small.

Within such approximations, the first term in (\ref{eq:flux_poisson}) is
\begin{eqnarray}
\label{eq:flux_poisson_simpl} & & \displaystyle \sum_{n,m} \nonumber
 \mathcal{P}_{\mu}(n) ~n f(n+m) \mathcal{P}_{\mu+\varepsilon}(m) =
\nonumber \\
& = & \displaystyle \sum_{n,m} \frac{e^{-\mu} \mu^n}{n!} n f(n+m)
\frac{e^{-(\mu + \varepsilon)} (\mu + \varepsilon)^m}{m!} =
\nonumber \\
& \simeq & \displaystyle \sum_{n,m} \frac{e^{-\mu} \mu^n}{n!} n
f(n+m) \frac{e^{-\mu}\mu ^m}{m!} (1 - \varepsilon) \left( 1 +
\frac{\varepsilon}{\mu} m \right) =
\nonumber \\
& \simeq & \displaystyle \sum_{n,m} \mathcal{P}_{\mu}(n)
\mathcal{P}_{\mu}(m) n f(n+m)\left[1+ \left( \frac{m}{\mu}-1
\right) \varepsilon \right]
\end{eqnarray}
where we neglected orders $\varepsilon^2$.

The second term in Eq.~\ref{eq:flux_poisson} can be rewritten
analogously as:
\begin{eqnarray}
\label{eq:flux_poisson_simpl2} & & \nonumber  \displaystyle
\sum_{n,m} \mathcal{P}_{\mu+\varepsilon}(m)~ m f(n+m) \mathcal{P}_{\mu}(n) =
\nonumber
\\ & \simeq & \displaystyle \sum_{n,m} \mathcal{P}_{\mu}(m)
\mathcal{P}_{\mu}(n) m f(n+m)\left[1+ \left( \frac{m}{\mu}-1
\right) \varepsilon \right]
\end{eqnarray}

By subtracting Eq.~\ref{eq:flux_poisson_simpl} and
Eq.~\ref{eq:flux_poisson_simpl2}:
\begin{eqnarray}
\displaystyle \sum_{n,m} \mathcal{P}_{\mu}(m) \mathcal{P}_{\mu}(n)
f(n+m) \frac{\varepsilon}{\mu}
 \left( m n - m^2 \right).
\end{eqnarray}

Notice that, due the symmetric sum, we can write $\left( m n - m^2
\right)$ as $- \frac{(m-n)^2}{2}$. Therefore we obtain:
\begin{equation}
\label{eq:flux_poisson_fin}
- \varepsilon \displaystyle \sum_{n,m} \frac{e^{-2 \mu} \mu^{n+m}}{n! m!} f(n+m)
\frac{1}{2\mu} \left( m-n \right)^2 =\Phi(i,i+1),
\end{equation}
to be compared with Eq.~\ref{eq:flux}. The analogy is actually
straight as $\varepsilon$ is just ${\partial n}/{\partial x}$.
Interestingly, from Eq.~\ref{eq:flux_poisson_fin} we obtain that
the approximate diffusivity ${\tilde D} ~$ is
\begin{equation}
{\tilde D}(\mu) = \displaystyle \sum_{n,m} \frac{e^{-2 \mu} \mu^{n+m}}{n!
m!} f(n+m) \frac{\left( m-n \right)^2 }{2 \mu},
\end{equation}
confirming that ${\tilde D}$ only depends on $\mu$. However, we recall
that this is still a local relation, since we
posed $\mu_i=\mu$.

Provided that $m$ and $n$ are sufficiently large, a simple
analytic relationship between ${\tilde D}(\mu)$ and $f(n+m)$ is obtained
approximating the Poissonian by a
Gaussian distribution ($\mathcal{P}_{\mu}(n) \simeq
\frac{1}{\sqrt{2 \pi \mu}} \;
e^{-\frac{(n - \mu)^2}{2 \mu}}$) and adopting a continuous
picture:
\begin{eqnarray}
\label{eq:diffusivity}
&  &     {\tilde D}(\mu)   \simeq  \\
& & - \int_{0}^{\infty}
\int_{0}^{\infty} \frac{ e^{-\frac{(x - \mu)^2}{2 \mu}}} {\sqrt{2
\pi \mu}} \; \frac{e^{-\frac{(y - \mu)^2}{2 \mu}}}{\sqrt{2 \pi
\mu}} \; \frac{\left( x-y \right)^2}{2 \mu} \; f(x+y) dx dy.\nonumber
\end{eqnarray}
Now, the variable change: $x+y = \xi$ and $x-y = \eta$ gives:
\begin{eqnarray}
\label{eq:diffusivity2}
& & {\tilde D}(\mu)   \simeq   \nonumber \\
& &  - \int_{0}^{\infty}\int_{-\infty}^{\infty} {e^{-\left[
\frac{(\xi + \eta)^2}{2} + 2 \mu^2 + \frac{(\xi - \eta)^2}{2} - 2
\mu \xi
\right]/(2 \mu)}}  \eta^2 \; f(\xi) \frac { d\xi \; d\eta} {8 \pi\mu^2}\;
 \nonumber \\
& & = \frac{1}{2 \sqrt{\mu \pi} e^{\mu}} \int_{0}^{\infty} f(\xi)
e^{-\xi^2 / 4\mu +\xi} d\xi.
\end{eqnarray}

It is now suitable to explicit the transition probability
$f(n_i,n_{i+1})$. We recall that it should depend on the link
energy and vanish for $E_i=N_c~$. We therefore define
\begin{equation}
\label{eq:transition}f(\xi) = f(x+y) = \frac{1}{2} \left[
1-e^{(\xi/2 - N_c)^2} \right]~,
\end{equation}
and this expression will be used in numerical simulations. Once
inserted in Eq. \ref{eq:diffusivity2}, such a transition probability
provides a close form for the diffusivity:
\begin{equation}
\label{eq:diffusivity_fin} {\tilde D}(\mu) = \frac{1}{2} \left[ 1 -
\frac{1}{\sqrt{\mu + 1}} \; e^{-(\mu - N_c)^2/(\mu +1)} \right] ~.
\end{equation}
Notice that ${\tilde D}(\mu)$ is qualitatively analogous to
$f(\mu)$ (Eq. \ref{eq:transition}) but it does not vanish for $\mu
= N_c$. Consequently, $\frac{\partial n}{ \partial x}$ does not
diverge and
 $N_c$ is not a true critical density, but only a value where
the dynamics presents a slowing down.
The absence of critical phenomena was already foreseeable
by considering that at the equilibrium the distribution is regular
for any value of $\mu$ and $N_c$.

\section{Numerical Results}
\label{sec:numerical}

Before discussing the results obtained from simulations, we briefly
resume how numerical experiments work. A system, characterized
by parameters $(L,N_1,N_L,N_c)$, is prepared with a uniform
distribution of particles. At each unit time, a randomly extracted walker
hops according to the transition probability given by Eq.
\ref{eq:transition}. Then, the system evolves along this rule,
generating time series for each observable. Such series evidence
how the system looses the memory of the initial distribution,
eventually reaching a steady state that depends only on the system
parameters.
To any observable we can associate its mean value
$\langle \cdot \rangle$ and fluctuation $F(\cdot)~$. Such
quantities are evaluated by considering a large number ($\sim
10^5$) of different independent realizations.

\subsection{Diffusivity}

\begin{figure}
\resizebox{0.9\columnwidth}{!}{
\includegraphics{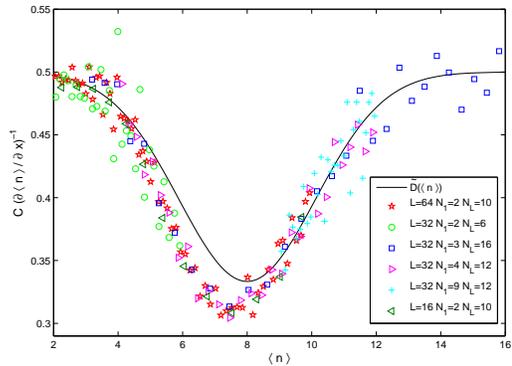}}
\caption{Simulations for $N_c=8$. Data points (symbols) and
theoretical results  $\tilde D(\langle n\rangle)$ (line) for $C (
{\partial \langle n \rangle}/{ \partial x})^{-1}$; different
symbols refer to systems of different sizes and boundary
conditions, as explained by the legend. The constant $C$ has been
fixed for different data in order to obtain the curve
collapse. The theoretical curve is given by
Eq.~\ref{eq:diffusivity_fin}} \label{fig:one}
\end{figure}

\begin{figure}
\resizebox{0.9\columnwidth}{!}{
\includegraphics{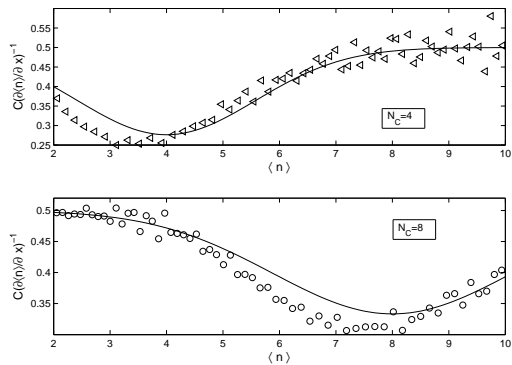}}
\caption{Data points (symbols) and theoretical predictions (lines)
for the diffusivity as a function of the particle density for two values
of the parameter $N_c$. The numerical data are obtained for systems of size
$L=64$, with $N_1=2$ and $N_2=10$.} \label{fig:fit}
\end{figure}

Let us first verify that the system can be described
by means of a Fourier like equation with a density dependent
diffusivity. For this purpose we show that equation
(\ref{eq:stationary_flow}) holds for different boundary conditions
and system sizes. In Fig.~\ref{fig:one} we have plotted the value
of the discrete derivative $\left (  {\partial \langle n
\rangle}/{ \partial x} \right)^{-1}$ (multiplied by a suitable
constant)  as a function of $\langle n\rangle~$. Different markers
correspond to different boundary conditions and sizes. The data
collapse is very good, confirming the general picture of a density
dependent diffusivity. The continuous line represents the
theoretical prediction given by Eq.~\ref{eq:diffusivity_fin}. We
recall that Eq.~\ref{eq:diffusivity_fin} was derived under the
assumption that the global distribution $P(\{ n \})$ is factorized
into $L$ Poissonians $\mathcal{P}_{\mu_i}$, moreover we
approximated such Poissonians with Gaussian distributions. Within
such approximations, the non perfect coincidence between the
analytic and the experimental minima can be considered as natural.
The calculated expression ${\tilde D}(\mu)$ provides therefore a
good estimate of the diffusivity, even for small sizes and
occupation numbers. In Figure \ref{fig:fit}, such a good agreement
is evidenced for different values of $N_c$.

Looking at the well collapsed minima, numerical  experiments
confirm that the diffusivity is strictly positive: consequently,
we do not expect any critical effects. More precisely, the site
$i_c$ (i.e. the site approximately corresponding to the density
$N_c$) does not give rise to any singularities in the observables
describing the system. Further evidences are given in the
next sections.

\subsection{Poissonian}

\begin{figure}
\resizebox{0.9\columnwidth}{!}{
\includegraphics{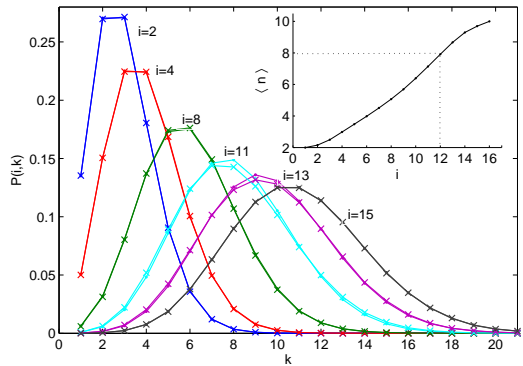}}
\caption{Occupation number distributions $P(i,k)$ for $L=16$,
$N_1=2$, $N_L=10$ and $N_c=8$. Different colours correspond to
different sites $i$ and different averages $\mu_i$. The Poissonian
distributions ($\bullet$) $\mathcal{P}_{\mu_i}(k)$ for a close
system with density $\mu_i$, are compared to the curves
($\times$), describing the open system with  gradient; lines are
guide to the eyes. The inset shows the average number $\langle n
\rangle$ as a function of the site index $i$: $\langle n \rangle
=8 $ approximately corresponds to $i=12$. The largest
discrepancies between $P(i,k)$ and $\mathcal{P}_{\mu_i}(k)$ occur
for sites corresponding to occupation numbers around $N_c$.}
\label{fig:pois}
\end{figure}

We study now the probability distribution for
the particle population in the presence of a gradient.
As proved in Section \ref{sec:factorization}, for a closed system
the occupation number is a stochastic variable described by a
Poissonian distribution. For an open system with a steady particle flow,
the average occupation number is site dependent and its distribution
is no more rigorously described by such a Poissonian.
However, we expect that the differences between the exact
 distribution $P(\{ n \})$ and $ \prod
\mathcal{P}_{\mu}$ is small when the  size $L$ is large with
respect to the particle gradient $N_L - N_1$. We also expect
that, the farthest from $i_c$, the better the approximation
works.
Indeed, these facts are confirmed by numerical simulations. For
example, in Fig.~\ref{fig:pois} we show the probability
distributions for the occupation number relevant to several sites
of an open system of size $L=16$, with $N_1=2, N_L=10$ and
$N_c=8$. Each site $i$  considered (depicted in a different colour)
corresponds to an average occupation number $\mu_i$, and the
pertaining distribution is compared with the Poissonian of average
$\mu_i$, i.e. the distribution for a close system made up of $N =
L \times \mu_i$ particles. As expected, near the borders the two
distributions perfectly overlap; in general, their difference is
small away from $i_c$ (which is about $12$, as pointed out in
the inset). Otherwise stated, the difference is appreciable in a
region around $i_c$. Moreover, experiments show that such a region
sensibly shrinks as the size gets larger.

\subsection{Fluctuations}

Let us study the occupation number
$\langle n_i \rangle$ and its fluctuations $F(n_i)$ as functions
of the system parameters $(L,N_1,N_L,N_c)$. The interest
in this point depends on the possibility, for interacting systems,
of an increase in fluctuations induced by the presence of a gradient,
as found e.g. in \cite{acv,dhar}.
More precisely, the system considered in \cite{acv} was an Ising model
on a cylindrical lattice, in contact with two thermostats at
temperatures $T_1$ and $T_2$. In the presence of a heat flow ($T_1 \neq T_2$),
it was found that fluctuations (generically denoted as $F$)
relevant to several observables all satisfy the following
$$
\Delta F \equiv  F_{\mathrm{flow}} - F_{\mathrm{no-flow}} \geq 0~,
$$
where $F_{\mathrm{flow}}$ and $F_{\mathrm{no-flow}}$ represent
the fluctuations in the system with and without flow, respectively.
Moreover, such difference is especially important in a domain
$\delta E$ of energies around
$E_c$, the critical energy of the Ising system at equilibrium.

Clearly, analogous comparisons can be carried out for our RW
model, by considering fluctuations in the occupation number for
open steady systems in the absence  ($N_1 = N_L$) or presence
($N_1 \neq N_L$) of a particle flow. Notice that, apart from
negligible border effects, the former case is well described by
the close system of Section \ref{sec:factorization} where the exact value
of the fluctuations can be evaluated from the Poissonian
distribution (i.e.  $F_{\mathrm{no-flow}}=\langle n \rangle $),
allowing for an accurate insight into the problem.

\begin{figure}
\resizebox{0.9\columnwidth}{!}{
\includegraphics{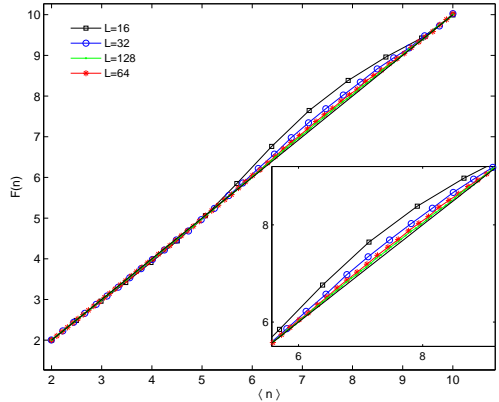}}
\caption{Fluctuations in the local occupation number $F(n_i)$ as a
function of $\langle n_i \rangle$. Different sizes are depicted, a
shown by the legend. $N_1=2, N_2=10$.} \label{fig:flutt_size}
\end{figure}

\begin{figure}
\resizebox{0.9\columnwidth}{!}{
\includegraphics{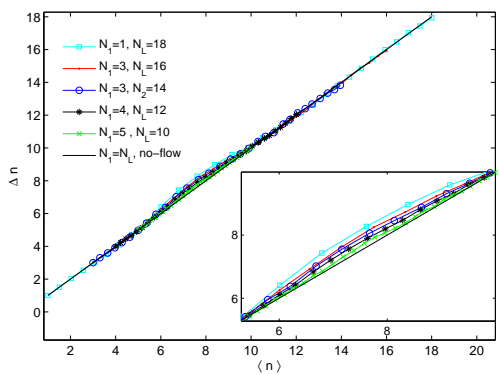}}
\caption{Fluctuations in the local occupation number $F(n_i)$ as a
function of $\langle n_i \rangle$. Different gradients are
depicted, as shown by the legend. $L=32$.} \label{fig:flutt_grad}
\end{figure}

Figures \ref{fig:flutt_size} and \ref{fig:flutt_grad} show $\Delta
F$ as a function of $\langle n \rangle$ for different choices of
the system parameters. Actually,  when a steady particle flow
is established, ({\it i}) $\Delta F \geq 0$, ({\it ii}) at a density
$\langle n \rangle \simeq N_c$  deviations from the equilibrium
values are stronger, and ({\it iii}) the overall effect is more
significant in a region $\delta N \equiv (N_c - \delta_1, N_c +
\delta_2)$. The analogy with the Ising system is notable.

Figures \ref{fig:flutt_size} and \ref{fig:flutt_grad} also
highlight the role of the size and of the density gradient: by
enlarging $L$ or by reducing $N_L - N_1$, the region
$\delta N$ gets smaller and the discrepancy $\Delta F$ shrinks.
Again, this effect is consistent with the results found for the
ferromagnetic model undergoing a steady heat flow.

Thus, both the ferromagnetic and the RW systems exhibit
an increase in fluctuations when a gradient is established.
However, such a gradient is not a sufficient cause:
sites have to interact so that a dynamical slowing down takes place
for a value of the parameter (energy and particle density
respectively).
For the Ising model, such a special value of the energy corresponds
to the critical point. As for  RW, it is the parameter $N_c$ in the
interaction defined by Eq. \ref{eq:transition}.

Now, Ising and RW systems display an important difference:
while for $L \to \infty$ the former exhibits
a true critical point, the latter is never critical, since $N_c$
does not correspond to any singular behaviour. In particular,
$\frac{\partial \langle n \rangle}{\partial x}$
(to be related to the diffusivity) is always finite.

We can conclude that the similar increase in fluctuations, observed
in both systems, is not really due to criticality but only to
a dynamical slowing down.
This solves the apparent contrast remarked in the Ising case
between the mesoscopic nature of the effect and the thermodynamic
nature of the critical point.

\subsection{Scaling}

\begin{figure}
\resizebox{0.9\columnwidth}{!}{
\includegraphics{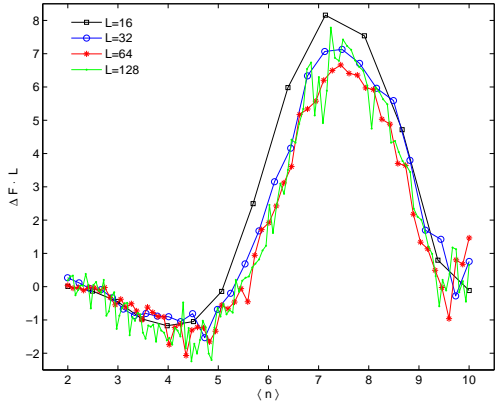}}
\caption{$\Delta F $ for
systems of different sizes and equal boundary
conditions, i.e. $N_1=2$ and $N_2=10$. The peak
corresponds to $N_c=8$. Different sizes are shown with different
colours and symbols. Lines are guide for
the eye.} \label{fig:scal}
\end{figure}

Previous observations can be stressed by looking directly at the
scaling behaviour. In general, as well known, fluctuations are
sensitive to criticality. The functional dependence on the size $L$,
for instance, is  different at the critical point. This
happened indeed in the Ising model at $E_c$ \cite{acv}. Now, since
in the present case $N_c$ is not a genuine critical point, as $L$
grows we expect, conversely, a regular behaviour for all occupation
numbers, those approaching $N_c$ included. Such an expectation is
confirmed by experiments reported in Figure \ref{fig:scal}, where
$\Delta F (\langle n \rangle)$ goes to zero as $L$ gets larger
according to the scaling law: $\Delta F \sim 1/L$, with an
excellent data collapse. The only relevant effect appearing at
$N_c$ regards the amplitude of $\Delta F $, which has a maximum,
not a distinct scaling law.

\subsection{Spectral features}
Time series can provide further information on the pseudo criticality around
$N_c$. It is worth summarizing how this kind of analysis is performed.\\
For a given site $i$, consider the sequence of occupation numbers:
$$
\{ n_i(t_1), n_i(t_2), ..., n_i(t_n) \},
$$
where $t_1, t_2, ..., t_n$ are instants of time after the onset of a steady
state. This definition may be implemented by introducing a time
delay $\Delta t$, and the corresponding time series:
$$
\{ n_i(t_1), n_i(t_1 + \Delta t),  n_i(t_1 + 2 \Delta t), ..., n_i(t_1 +
k \Delta t),... \}.
$$
Clearly, the delay or sampling parameter $\Delta t$ plays an important role
in experiments, since time correlations are strong in our model, and
a sufficiently long interval is needed before a given configuration
significantly changes.

Then, by Fast Fourier Transform, from time series we get power spectra $S(f)$,
i.e. the square of absolute Fourier transform amplitudes in the frequency domain.
The so-called colour exponent  $\alpha$ describes the possible power-law
decay $S(f) \sim  f^{\alpha}$ of the spectrum. It can be obtained as
the angular coefficient of the linear fit in the log-log plot of $S(f)$ vs. $f$.

Of course, a special value is  $\alpha = 0$ associated to white
noise, decorrelation, randomness. Values  $ \alpha <0 $ qualify
coloured noises corresponding to different types of
temporal correlations. In particular, $\alpha = -1$ (``pink''  or
$1/f$ noise) implies an extremely slow decay of correlations.

As it is well known, a general dynamic theory of coloured noise is still
lacking (see however \cite{weissman}). The old suggestion of Van der Ziegle,
getting $1/f$ noise from the superposition of independent Poisson processes
\cite {ziegle,jensen}, could be of some interest in our case.

\begin{figure}
\resizebox{0.9\columnwidth}{!}{
\includegraphics{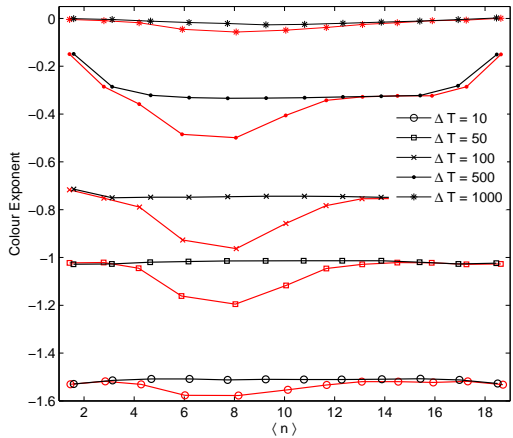}}
\caption{Color exponent for a system of interacting (red) and non
interacting (black) particles, both sized $L=16$ and with $N_1=
0.5, N_L =20.0$; for the former $N_c=8.0.$ Different time delays
$\Delta t$ are depicted, as shown in the legend.}
\label{fig:color}
\end{figure}

In simulations, for every $\Delta t$, exponents are evaluated and
averaged over several runs starting from different initial
conditions, up to stabilization. The dependence of the noise on
interaction is shown in Fig.~\ref{fig:color}, where exponents
$\alpha$ versus $\langle n \rangle$ are plotted for analogous
systems made up of interacting and non-interacting (i.e.
isotropically diffusing) particles; several choices of $\Delta t$
are also considered. Systems of larger sizes, not reported here,
display qualitatively alike outlines.

The presence of interaction clearly  shifts $\alpha$ towards
smaller values, but only in a domain around $N_c$ largely overlapping
the previously considered domain  $\delta N$.
Indeed, $\alpha$ is minimum for $n \simeq N_c$. This
confirms that around $N_c$ correlations are stronger and the
dynamics is slowed down. Moreover, the comparison with the
non-interacting system allows to figure out the background
time-correlation which, as expected, decreases as $\Delta t$ is
larger, up to $\alpha =0$. \\
A less trivial point has to be underlined: the difference between
interacting and non-interacting case is not a monotonic function
of $\Delta t$ and, therefore, neither of $\alpha$ itself.
Actually, there exists an optimal time delay $\tilde{\Delta t}$
maximizing the effect of interaction over time correlation.
Such a special delay $\tilde{\Delta t}$ seems difficult to be
interpreted, and could suggest that there exist further
aspects, besides the superposition of
Poisson processes, characterizing the noise in our model.

\section{Conclusions and perspectives }
\label{conc}

We have introduced a system of interacting random walkers modelling
several situations where diffusivity depends on the density of the
diffusing entities. Our model constitutes a special
case of misanthrope processes, characterized by a totalistic hopping
amplitude. We have shown analytically that the equilibrium distribution
is factorized into a product of Poissonian functions.

Despite a simple equilibrium behaviour, far from equilibrium  the study
of the system presents some non-trivial tasks.
Indeed, in principle, local properties such as correlations and factorizability
are deeply influenced by non equilibrium.
In our case, we succesfully explored the reliability of a perturbative
approach to the problem. Since, on one hand, the equilibrium properties
are analytically known and, on the other, interaction still plays a relevant
role, our model can be considered as an interesting benchmark for the study
of non-equilibrium effects.
By recovering analogous results in magnetic models, we have clarified
some open problems, concluding, in particular, that a density dependent
diffusivity  and the presence of a dynamical slowing down are necessary
premises to the amplification of fluctuations, while, on the contrary,
the existence of a true critical point may cooperate but it is not
strictly necessary.

Finally, even if the passage to regular lattices of higher dimensionality
does not present any substantial novelty, we evidenced a possible deep
interplay between time scaling, substrate topology and local interactions.
Consequently, there still exist in this context a number of
geometro-dynamical aspects  which have to be developed,
understood and classified. The non trivial influence of
interaction on the noise can be considered as a notable hint
for the relevance of the matter.

\end{document}